# Embedded Waveguide Structures Fabricated in Porous Silicon by means of Lithography

**Alexander Kellarev, Shlomo Ruschin**
*Department of Physical Electronics, Faculty of Engineering,*
*Tel-Aviv University, Tel-Aviv 6997801, Israel*
kellarev@tauex.tau.ac.il; ruschin@tauex.tau.ac.il

## Abstract

We report the fabrication of diverse light guiding structures embedded in oxidised porous silicon (OPS) by means of patterned pore filling through a single photoresist mask. The pores of OPS were filled with KDP salt from unsaturated solution applied on the sample's surface. Photolithography was used to create the mask. The created structures include straight, bent and tapered waveguides of different widths, and Y-splitters. Light guiding was demonstrated and characterised at wavelength of 1.064 μm.

## Introduction

Current technologies allow fabrication of diverse types of waveguides on different substrate materials. These types include embedded, rib, ridge waveguides and others. [1] Si, InP, $LiNbO_3$ and glass are customary examples of the substrate materials used for the waveguide creation. Specifically, silicon is an attractive material for photonic applications due to the existing large infrastructure for processing it, capable of mass-production of silicon-based circuits. The technologies of waveguide fabrication are based on the creation of light-guiding regions of high refractive index and involve various processes in accordance with the waveguide type and materials used.

Patterned pore filling in porous medium, reported here, is a novel method of waveguide fabrication, which is aimed to be intuitive, relatively simple and cost effective. The main idea of this method is that filling the pores of a porous medium results in increase of its effective refractive index, and it can be performed locally.

Application of the patterned pore filling for creation of optical waveguides was reported in a previous publication, where it was accomplished by means of a material printer. [2] The waveguides were printed with aqueous KDP ($KH_2PO_4$) solution on oxidised porous silicon (OPS) substrates. The width of the printed waveguides was around 20 μm. Light guiding was demonstrated and propagation losses were measured at wavelengths 1.064 and 1.550 μm. The values of the propagation loss at these wavelengths were 2.4 dB/cm and below 0.5 dB/cm respectively.

Being very intuitive and relatively simple process, printing of optical waveguides is currently limited by the existing ink deposition technologies, which allow minimum printed feature size of order of tens of micrometres. Smaller waveguide width would be desirable in order to increase photonic integration density. This limitation can be overcome by implementing lithography with patterned pore filling.

In the presented work we used the method of patterned pore filling with KDP salt for the fabrication of more complicated light guiding structures, which included tapered and bent waveguides and Y-splitters, in addition to the straight waveguides. The patterns of the structures were created as open areas in a photoresist mask, through which the applied KDP solution was able to fill the pores in an OPS substrate. We used lithography to create the mask on top of the substrate, and removed the mask after the pore filling process. Lithography allowed fabrication of narrow waveguides, as was necessary for single-mode operation at the tested wavelengths. Light propagation through the created structures was characterised by means of imaging at a wavelength of 1.064 µm. In the following sections we present a brief theoretical background of the effect of patterned pore filling, descriptions of the fabrication and characterisation methods, and results followed by a discussion.

## Theoretical background

Porous silicon (PS) was discovered by A. Uhlir in 1956. [3] This material possesses extremely large surface area and very high ability to absorb other substances in its pores. Oxidation converts PS into oxidised porous silicon (OPS), which is hydrophilic. [4] PS and OPS behave as optical metamaterials regarding their interaction with light of wavelength much larger than the pore diameter. These properties make PS and OPS very attractive for various applications, including sensors with optical readout and optical integrated circuits. [5]

PS is produced by means of electrochemical etching of a Si wafer, resulting in the formation of a porous layer on its surface. [4] The pore diameter and the layer thickness are determined by the wafer resistivity, current density and etching time. Etching at different current densities allows the creation of several porous layers with different pore diameters on the same wafer. OPS can then be obtained by means of thermal oxidation of PS, which provides environmental stabilisation and hydrophilicity.

When treated as optical metamaterial, the effective index of refraction of PS and OPS can be described by means of a generalised Bruggeman model. [6], [7] According to this model, filling voids in a porous material will alter its optical properties. Specifically, insertion of any material with refractive index greater than one into the pores of PS or OPS will increase its effective refractive index. A simple way to fill the pores in OPS is applying an aqueous salt solution on the OPS surface, followed by drying. Experimental study of the pore filling of OPS with salts was reported in our previous publication. [8] It was demonstrated that the increase of the refractive index of the OPS could be controlled by the solution concentration and can be spatially confined by applying patterned pore filling.

Light guiding requires cladding regions of low refractive index adjacent to the core with high refractive index. This requirement implies using a substrate with more than one layer, in which the porous layer containing the core is surrounded by layers with lower refractive indices. In our case we chose a two-porous-layer substrate, where the first (top) porous layer has refractive index higher than that of the second (bottom) porous layer. Air above the top layer acts as the additional cladding layer. The required difference in the refractive indices can be achieved by etching the Si wafer at different current densities, such that pores in the top layer are of smaller diameter than pores in the bottom layer. Patterned application of a salt solution to a such substrate will result in creation of regions of rectangular shape in the top layer, where the filling material tends to remain due to stronger capillary effect in the smaller pores. A schematic drawing of a cross-section of the desired waveguide structure is shown in Figure 1.

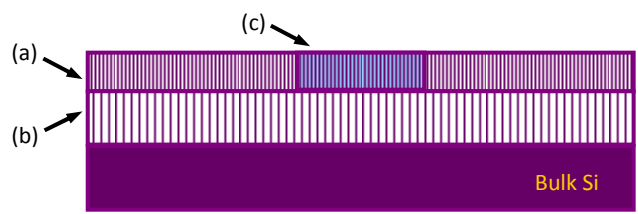

Figure 1: Schematic drawing of cross-section of an embedded waveguide in OPS; (a) – high refractive index porous layer, (b) – low refractive index porous layer, (c) – filled region.

## Fabrication

We used a (100)-cut p-type Si wafer with resistivity of 0.02..0.03 Ohm·cm to create a two-layer sample by means of electrochemical etching at different current densities. The top layer was etched at current density of 30 mA/cm$^2$. The bottom layer was etched at current density of 50 mA/cm$^2$. The etching times were 140 s and 200 s correspondingly. The porous region formed by the etching was of circular shape of diameter 15 mm in a sample of size 20 mm by 20 mm. The thickness of the porous layers was 3.4 μm (top) and 6.4 μm (bottom), measured from microscope images of the cross-section of the sample. After the etching the samples were thermally oxidised by means of a two-step process: pre-oxidation at 300⁰C followed by oxidation at 900⁰C for 2 hours. Refractive indices of the OPS layers were estimated from measurements of reflection spectra of single-layer samples etched at the same current densities. At wavelength of 1.064 μm the refractive indices were 1.352 and 1.204 for the top and the bottom layers respectively, providing the difference of 0.15.

Photolithography was used to delineate a photoresist mask on the surface of the sample. The standard lithography process was tuned in order to minimise the possibility of photoresist entering the pores. The purpose of the photoresist was to protect the pores from the liquid during the pore filling step, allowing filling the pores only in the open regions as defined by the mask pattern. The mask, shown in Figure 2, was designed to create the following structures:

- narrow waveguide of width 3 μm
- wide waveguide of width 30 μm
- tapered waveguide of width 3 μm with entrance width of 30 μm
- narrow waveguide of width 3 μm with S-bend
- Y-splitter with narrow waveguides of width 3 μm
- Y-splitter with wide waveguides of width 30 μm
- tapered Y-splitter of width 3 μm with entrance width of 30 μm.

The radius of curvature in S-bends and Y-splitters was 2 mm. The Y-splitters were designed symmetric aimed to a split ratio of 50/50. Spacing between the branches of the 3-μm Y-splitters was 251 μm. Spacing between the branches of the 30-μm Y-splitters was 224.1 μm. Some additional circular and rectangular shapes were made in the mask for alignment and testing purposes.

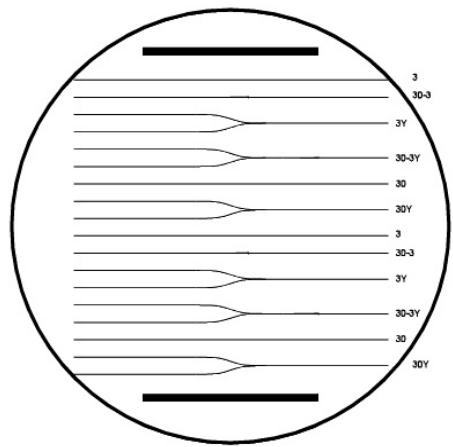

Figure 2: Lithography mask, the black lines designate open areas.

Pore filling was achieved by application of aqueous KDP solution of concentration 5.1% (by weight) on the sample's surface for 10 s, followed by drying on a P7600 spinner from Speciality Coating Systems. The dry KDP salt was purchased from Sigma-Aldrich. The pore filling resulted in increase in the refractive index of the top porous layer of 0.04 in the filled regions. The change in the refractive index was obtained from measurements of optical thickness of single-layer OPS samples before and after the pore filling, as described in [8]. Following the pore filling process, the photoresist was removed with acetone. Microscope image of the sample's surface showing part of the created pattern is seen in Figure 3. Each image in Figure 3 is a combination of several small images using Hugin software. In order to make the pattern visible the microscope was slightly defocused from the observed surface. The actual line width measured from the surface images was 24 µm for the lines designed as 30 µm and 2.9 µm for the lines designed as 3 µm.

Sides of the sample were cleaved to allow access to the ends of the layered structure. The length of the cut sample was 6.4 mm. Cross-section of the sample was examined in a confocal microscope. A region with the filled pores can be seen in an image of the cross-section in Figure 4. As in the pore filling performed by means of printing, reported previously, the KDP salt remained mostly in the top porous layer and the waveguide core region appears well confined. [2]

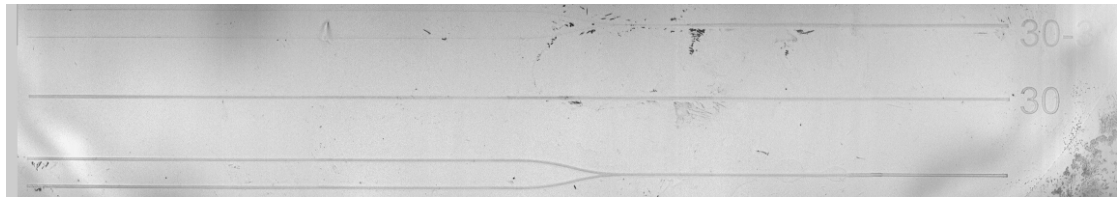

Figure 3: Images of the sample's surface.

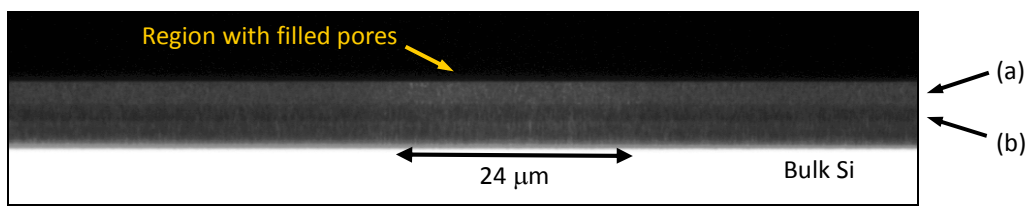

Figure 4: Cross-section of the sample showing a region with filled pores; thickness of the top layer is 3.4 µm, thickness of the bottom layer is 6.4 µm; (a) – top layer, (b) – bottom layer.

## Waveguide characterisation

Light propagation through the structures created in the sample was examined by means of end-coupling a 1.064-µm laser source through a lensed fibre and imaging the light scattered from the sample's surface and the transmitted light. The lensed fibre was aligned at the centre of the core of each structure. A Touptek UCMOS01300KMA camera with a 5x microscope objective was located above the sample in order to collect scattered light from the sample's surface. The transmitted light was imaged at the opposite side of the sample with a Duma Optronics BeamOn-HR camera through a 20x microscope objective.

## Results and discussion

The transmitted light and the light scattered from the surface were imaged for each of the structures created in the sample. Several images of the light transmitted through various structures are presented in Figure 5. As expected, propagation through the wide waveguides of width 24 µm results in wider spots in comparison to those of width 2.9 µm. From examination of the spatial distribution of the transmitted light it appears that the narrow waveguides are close to single-mode, while the wide waveguides transmit multiple modes. This comes in agreement theoretical and numerical estimations performed by means of Marcatili's method and Lumerical MODE software. [9] In some images the top porous layer appears slightly illuminated, implying the presence of scattering and radiative losses. This can be explained by remaining light scattering spots and by the fact that the top layer constitutes a planar waveguide itself. The other structures with a bend or a taper provided similar results.

The image data collected from the straight narrow waveguides were analysed in order to obtain spatial profiles of the transmitted light along the horizontal and vertical directions. By fitting the profiles extracted from the images to a Gaussian function in Matlab we obtained the size of the transmitted mode as 4.9 µm x 2.2 µm (X x Y). This result differs from the theoretical mode size of 2.2 µm x 2.1 µm, which was calculated by means of Lumerical MODE software. Analysis of the total light intensity transmitted through the branches of the Y-splitters delivered a split ratio of approximately 60/40 for both wide and narrow waveguides.

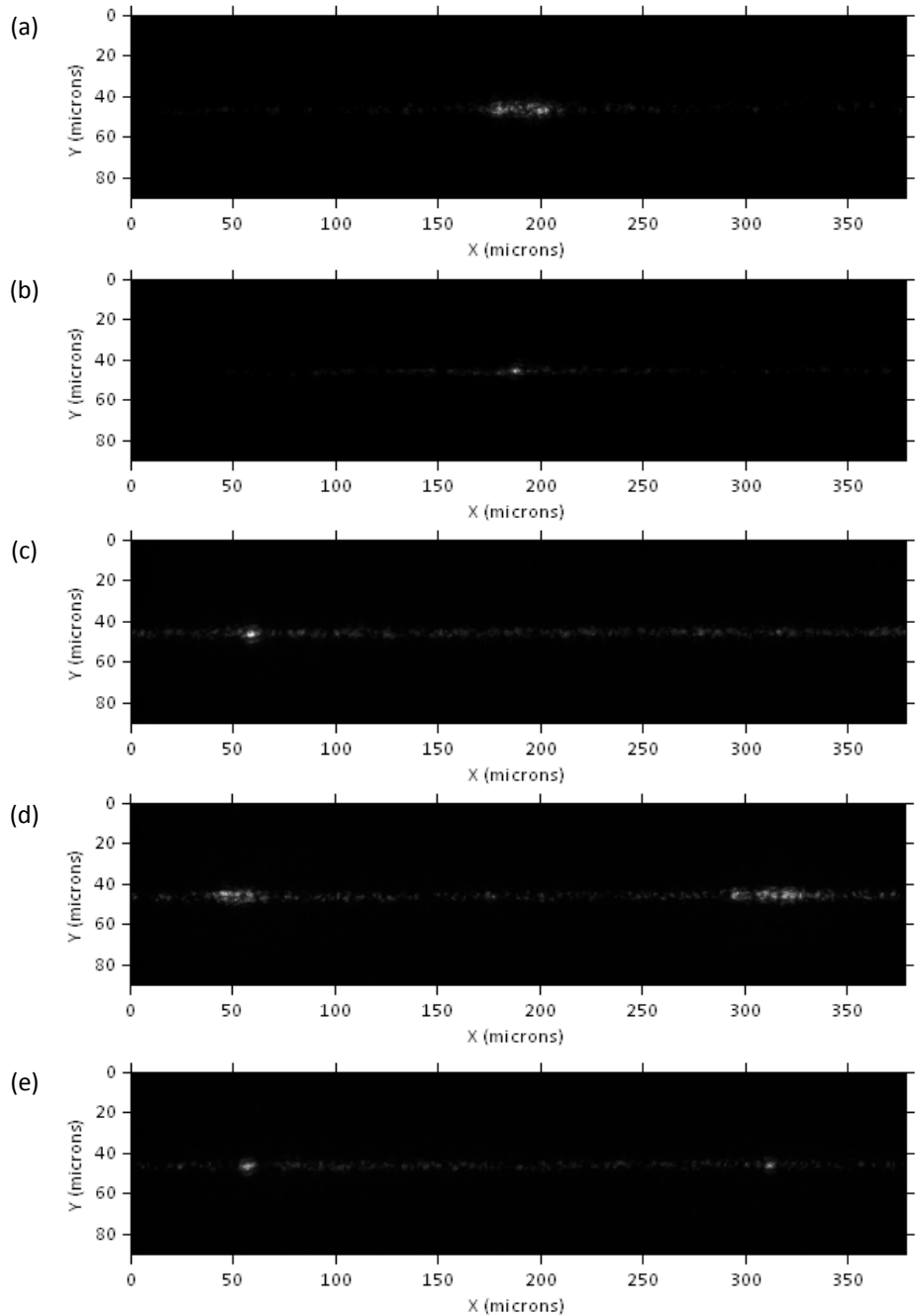

Figure 5: Images of the transmitted light; (a) - 24-μm waveguide; (b) - 2.9-μm waveguide; (c) - 2.9-μm waveguide with S-bend; (d) - 24-μm Y-splitter; (e) - 2.9-μm Y-splitter.

The top images of scattered light allowed analysing the lateral intensity profile of the guided light. The intensity profiles along the straight waveguides were used to obtain the propagation loss. [2], [10]–[14] The intensity was integrated across the waveguides for more stable results. The attenuation of the propagated light was determined by fitting the longitudinal profiles to an exponential function. The least absolute residuals method was used in Matlab software in order to obtain more robust results. The fitting was performed in a region excluding the entrance of the waveguides, where excessive scattering was observed. Example of the image of the light scattered along a 24-μm waveguide and the corresponding intensity profile are shown in Figure 6. The resulting propagation loss was 20±1 dB/cm for the wide waveguide of width 24-μm and 23±2 dB/cm for the narrow waveguide of width 2.9-μm.

The experimental results show that the measured lateral mode width differs from the theoretical one and the propagation losses are considerably high in comparison to similar waveguides fabricated in OPS by means of printing or other techniques. [2], [15], [16] Such differences can be explained by presence of residual materials used in the lithographic process, which eventually remained trapped in the porous layers after removal of the photoresist mask. The good correspondence between the estimated and the measured mode widths in the normal direction confirms the estimated refractive index values.

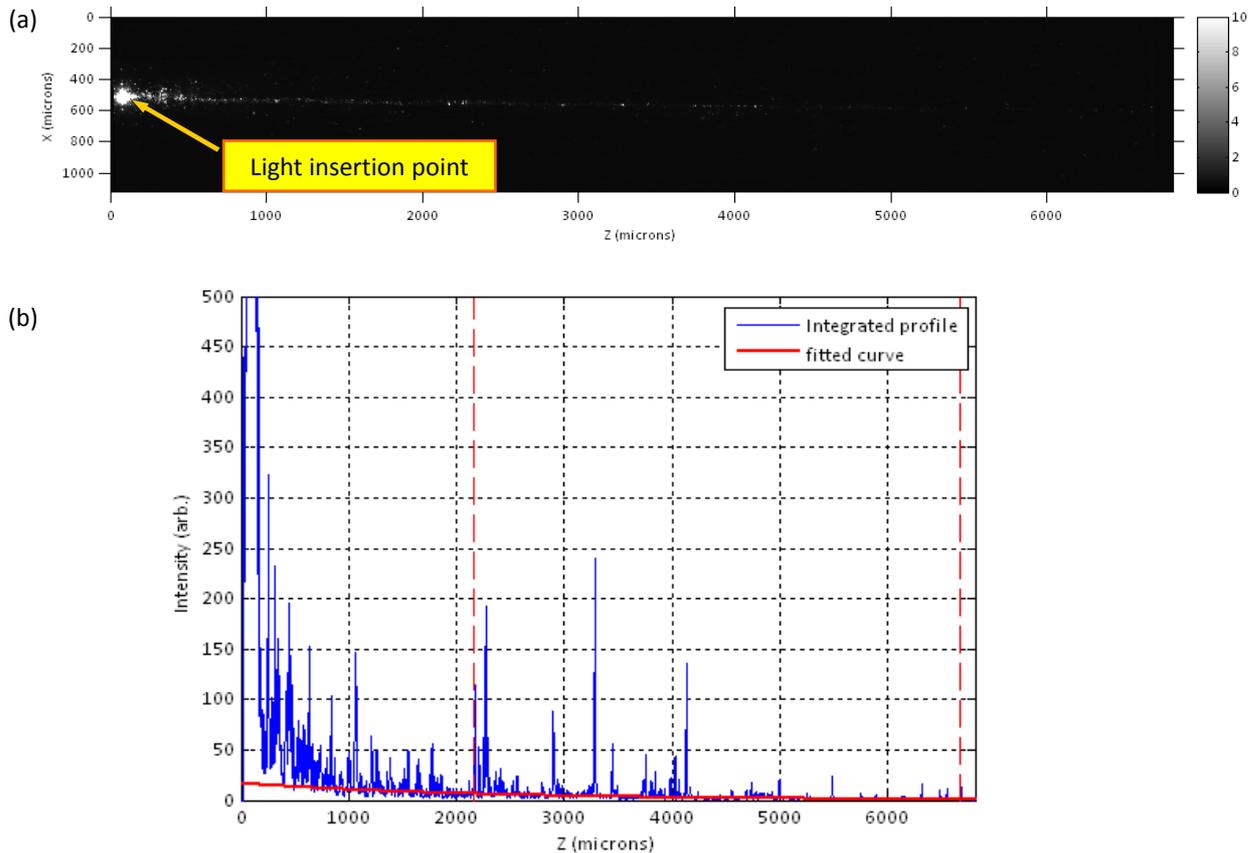

Figure 6: (a) - Image of the scattered light at λ=1.064 μm along the 24-μm waveguide; (b) - Intensity profile along the waveguide; vertical dashed lines designate the fitting region.

## Conclusion

We demonstrated the fabrication of various kinds of waveguides and splitters in oxidised porous silicon by means of pore filling. Patterned pore filling implemented with photolithography allowed creation of waveguides of widths down to 3 μm, which are significantly smaller in comparison to the printed waveguides reported previously by us. The fabrication process is simple and fast in comparison to current waveguide fabrication technologies requiring only one photolithographic mask. The measured propagation losses were relatively high, but we believe that they can be reduced by adjusting and improving the lithography process and the involved materials.